\documentclass[11pt,superscriptaddress,aps,prd,preprint,showpacs]{revtex4}

\usepackage[dvips]{graphicx}
\usepackage{amsfonts}
\usepackage{slashed}
\usepackage[utf8x]{inputenc}
\usepackage{amsmath}

\begin{document}

\title{Emergent gauge bosons and dynamical symmetry breaking in a four-fermion Lifshitz model}

\author{T. Mariz}
\author{R. Moreira}
\affiliation{Instituto de F\'\i sica, Universidade Federal de Alagoas,\\ 57072-900, Macei\'o, Alagoas, Brazil}
\email{tmariz,rmoreira@fis.ufal.br}

\author{A. Yu. Petrov}
\affiliation{Departamento de F\'{\i}sica, Universidade Federal da Para\'{\i}ba,\\
 Caixa Postal 5008, 58051-970, Jo\~ao Pessoa, Para\'{\i}ba, Brazil}
\email{petrov@fisica.ufpb.br}

\begin{abstract}
 We consider the four-fermion Lifshitz model, introduce in this model an auxiliary vector field, and generate an effective dynamics for this field. We explicitly demonstrate that within this dynamics, the effective bumblebee potential for the vector field naturally arises as a one-loop correction and allows for the dynamical breaking of the rotational symmetry.
\end{abstract}


\maketitle

The emergent dynamics is actually considered as one of the most appropriate explanations for the existence of various field theory models. Following this concept, the real physical theories arise from some fundamental models of interacting fermions as a result of spontaneous symmetry breaking or integration over some fields \cite{Bjorken}. The paradigmatic examples are Gross-Neveu and Thirring models, where the Lagrangians of (pseudo)scalar \cite{GN,NJL} and vector fields \cite{Thirring} respectively arise in this way. The importance of this concept is confirmed by the fact that actually the idea that gravity represents an emergent phenomenon as well, is intensively discussed \cite{Verlinde}, although it possesses a distinct motivation. 

Clearly, the concept of emergent dynamics is expected to be very natural within Horava-Lifshitz approach characterized by a strong asymmetry between space and time, so that the theory is invariant under anisotropic rescaling $x_i\to bx_i$, $t\to b^zt$, with $z$ being a critical exponent. As a result, the action of the theory involves two time derivatives and $2z$ space ones, or, for a spinor field, one time derivative and $z$ space ones. This approach has been formulated systematically in \cite{HL} within gravity context, where it was expected to yield renormalizable $(3+1)$-dimensional gravity. Further, this methodology has been generalized to other field theory models, which are frequently called the Lifshitz or Lifshitz-like models. Within this approach, there is no dynamics fixed {\it a priori}, since various models with different couplings and different critical exponents can be considered. Therefore, the emergent dynamics seems to be the most natural explanation for arising of different models for scalar and vector fields. Already in \cite{ourGN}, this mechanism was used to generate dynamics of a scalar field from the Lifshitz-like Gross-Neveu model. Important results in this context were also obtained in \cite{ourHLED}, where it was shown that the quadratic gauge invariant action for the vector field arises in the Lifshitz spinor theory with certain spinor-vector couplings. 

It should be noted that many issues related to perturbative aspects of various non-gravitational Lifshitz-like field theory models, including the effective potential and renormalization, have been studied up to now, see f.e. \cite{difpapers}. Therefore, it is natural to study the emergent dynamics of the vector field in the Lifshitz-like theory. To be more precise, in this paper we are going to study the rotational symmetry breaking generated by the emergent terms. This mechanism is known to be applied within the context of the Lorentz symmetry breaking, where it allowed to generate the bumblebee potential allowing for spontaneous symmetry breaking \cite{KosGra,Seifert} as a quantum correction \cite{bumb}. Thus, it is natural to generate the bumblebee potential within Lifshitz-like theories, with the only difference consists in the fact that in this case, this potential will generate dynamical breaking not of the Lorentz symmetry but of the spatial rotational symmetry. This is the aim of the present paper.

Within this text, we consider the Lifshitz four-fermion model, explicitly demonstrate that in this model, the vector field is naturally introduced as an auxiliary field, generate the one-loop bumblebee-like potential for this vector field, and demonstrate that this potential displays spontaneous breaking of the $O(N)$ rotational symmetry.

We start with formulating the $z=2n+1$ Lifshitz four-fermion model
\begin{equation}\label{L0}
{\cal L}_0 = \bar\psi (i\slashed{\partial}_0+(i\slashed{\partial}_i)^z-m^z)\psi - \frac{G_t}{2} (\bar\psi\gamma_0\psi)^2 - \frac{G_s}{2} (\bar\psi\gamma_i(i\slashed{\partial}_j)^{2n}\psi)^2,
\end{equation}
where $\slashed{\partial}_0=\partial_0\gamma^0$, $\slashed{\partial}_i=\partial_i\gamma^i$, $(\bar\psi\gamma_0\psi)^2=(\bar\psi\gamma_0\psi)(\bar\psi\gamma^0\psi)$, and
\begin{equation}
(\bar\psi\gamma_i(i\slashed{\partial}_j)^{2n}\psi)^2 = (\bar\psi\gamma_i(i\slashed{\partial}_j)^{2n}\psi)(\bar\psi\gamma^i(i\slashed{\partial}_k)^{2n}\psi).
\end{equation}
Unlike \cite{ourGN}, here we consider two different couplings allowing for introducing the auxiliary vector field, that is, for convenience, we distinguish its time and space components $A_0$ and $A_i$. Moreover, one of our four-fermion couplings involves derivatives. Taking into account $(i\slashed{\partial}_i)^{2n}=\Delta^n$, with $\Delta=-\partial_i\partial^i$, we get
\begin{eqnarray}\label{L02}
\mathcal{L}&=&\mathcal{L}_0+\frac{g_t^2}{2}\left(A_0-\frac{e}{g_t^2}\bar{\psi}\gamma_0\psi\right)^2+\frac{g_s^2}{2}\left(A_i-\frac{e}{g_s^2}\bar{\psi}\gamma_i\Delta^n\psi\right)^2\nonumber\\
&=&\frac{g_t^2}{2}A_0A^0+\frac{g_s^2}{2}A_iA^i+\bar{\psi}(i\slashed{\partial}_0+i\slashed{\partial}_i\Delta^n-e\slashed{A}_0-e\slashed{A}_i\Delta^n-m^z)\psi,
\end{eqnarray}
where $g_t^2=\frac{e^2}{G_t}$, $g_s^2=\frac{e^2}{G_s}$,  and we have introduced the notations $\slashed{A}_0=A_0\gamma^0$ and $\slashed{A}_i=A_i\gamma^i$. We note that the resulting Lagrangian cannot be expressed in terms of gauge covariant derivatives $D_{0,i}=\partial_{0,i}-ieA_{0,i}$. However, it is not necessary since our aim consists in generating the Lifshitz analogue of the bumblebee model, which, as it is known \cite{KosGra,Seifert}, does not possess the gauge symmetry. On the base of this Lagrangian, we can introduce the following generating functional
\begin{eqnarray}
Z(\bar \eta,\,\eta) &=& \int DA_\mu D\psi D\bar\psi\, e^{i\int d^4x({\cal L}+\bar\eta\psi+\bar\psi\eta)}\nonumber\\
&=& \int DA_\mu\, e^{i\int d^4x \left(\frac{g_t^2}{2}A_0A^0+\frac{g_s^2}{2}A_iA^i\right)} \int D\psi D\bar\psi\, e^{i\int d^4x(\bar\psi S^{-1}\psi+\bar\eta\psi+\bar\psi\eta)},
\end{eqnarray}
where $S^{-1}=i\slashed{\partial}_0+i\slashed{\partial}_i\Delta^n-e\slashed{A}_0-e\slashed{A}_i\Delta^n-m^z$ is the operator describing the quadratic action.  To integrate over the fermion fields, we make the shift $\psi\rightarrow \psi-S\eta$ and $\bar{\psi}\rightarrow \bar{\psi}-\bar{\eta}S$, so that we arrive at the transformation $\bar\psi S^{-1}\psi+\bar\eta\psi+\bar\psi\eta \rightarrow \bar\psi S^{-1}\psi-\bar\eta S \eta$. {As a result, we obtain
\begin{eqnarray}
Z(\bar \eta,\,\eta) &=& \int DA_\mu\, e^{i\int d^4x \left(\frac{g_t^2}{2}A_0A^0+\frac{g_s^2}{2}A_iA^i\right)} \int D\psi D\bar\psi\, e^{i\int d^4x(\bar\psi S^{-1}\psi-\bar\eta S \eta)}.
\end{eqnarray}
Finally, integrating over fermions, we find the result for the generating functional
\begin{equation}\label{GF}
Z(\bar \eta,\,\eta) = \int DA_\mu \exp\left(iS_\mathrm{eff}[A] - i\int d^4x\, \bar\eta\, S\, \eta \right),
\end{equation}
where the effective action is given by
\begin{equation}\label{Seff}
S_\mathrm{eff}[A] = \int d^4x\, \left(\frac{g_t^2}{2}A_0A^0+\frac{g_s^2}{2}A_iA^i\right) -i \mathrm{Tr} \ln(\slashed{p}_0+\slashed{p}_i(p_jp^j)^n-e\slashed{A}_0-e\slashed{A}_i(p_jp^j)^n-m^z).
\end{equation}
The $\mathrm{Tr}$ symbol stands for the trace over Dirac matrices as well as for the integration in momentum or coordinate spaces. The matrix trace can be easily calculated, so that for the effective potential, we get
\begin{equation}\label{Vef}
V_\mathrm{eff} = -\left(\frac{g_t^2}{2}A_0A^0+\frac{g_s^2}{2}A_iA^i\right) +i \mathrm{tr} \int\frac{d^4p}{(2\pi)^4}\, \ln(\slashed{p}_0+\slashed{p}_i(p_jp^j)^n-e\slashed{A}_0-e\slashed{A}_i(p_jp^j)^n-m^z).
\end{equation}

The next step is to investigate the gap equations, which in this case are given by
\begin{eqnarray}\label{gap0}
\frac{\partial V_{\mathrm{eff}}}{\partial A_0}\Big|_{A_\mu=\frac{a_\mu}{e}}&=&-\frac{g_t^2}{e}a^0-ie\Pi^0_z=0, \\
\frac{\partial V_{\mathrm{eff}}}{\partial A_i}\Bigr|_{A_\mu=\frac{a_\mu}{e}}&=&-\frac{g_s^2}{e}a^i-ie\Pi^i_z=0,
\end{eqnarray}
where
\begin{eqnarray}
\label{pi}
\Pi^0_z &=& \mathrm{tr} \int\frac{d^4p}{(2\pi)^4}\frac{1}{\slashed{p}_0+\slashed{p}_i(p_jp^j)^n-\slashed{a}_0-\slashed{a}_i(p_jp^j)^n-m^z}\gamma^0, \\
\Pi^i_z &=& \mathrm{tr} \int\frac{d^4p}{(2\pi)^4}\frac{1}{\slashed{p}_0+\slashed{p}_i(p_jp^j)^n-\slashed{a}_0-\slashed{a}_i(p_jp^j)^n-m^z}\gamma^i(p_jp^j)^n.
\end{eqnarray}
 We use the fact that, independently of nature and dimension of $q_0$ and $q_i$, for the signature $(+,-,-,-)$, we have
\begin{eqnarray}
\frac{1}{\gamma^0q_0+\gamma^iq_i+M}=\frac{\gamma^0q_0+\gamma^iq_i-M}{q^2_0-q_iq_i-M^2}.
\end{eqnarray}
Thus, we get
\begin{eqnarray}
\Pi^0_z &=& \mathrm{tr} \int\frac{dp_0d^d\vec{p}}{(2\pi)^4}\frac{\gamma^0(p_0-a_0)+\gamma^i(p_i-a_i)(p_kp^k)^n+m^z}{(p_0-a_0)^2+(p_l-a_l)(p^l-a^l)(p_jp^j)^{2n}-m^{2z}} 
\gamma^0, \\
\Pi^i_z &=& \mathrm{tr} \int\frac{dp_0d^d\vec{p}}{(2\pi)^4}\frac{\gamma^0(p_0-a_0)+\gamma^j(p_j-a_j)(p_kp^k)^n+m^z}{(p_0-a_0)^2+(p_l-a_l)(p^l-a^l)(p_cp^c)^{2n}-m^{2z}} \gamma^i(p_mp^m)^n.
\end{eqnarray}
Now, we can easily calculate the trace, so that we obtain
\begin{eqnarray}
\Pi^0_z &=& 4 \int\frac{dp_0d^d\vec{p}}{(2\pi)^4}\frac{(p_0-a_0)}
{(p_0-a_0)^2+(p_l-a_l)(p^l-a^l)(p_jp^j)^{2n}-m^{2z}} , \\
\Pi^i_z &=& -4\delta^{ij} \int\frac{dp_0d^d\vec{p}}{(2\pi)^4}\frac{(p_j-a_j)(p_kp^k)^{2n}}
{(p_0-a_0)^2+(p_c-a_c)(p^c-a^c)(p_lp^l)^{2n}-m^{2z}} .
\end{eqnarray}
Here, an essential difference between time and space components takes place. Indeed, let us do, in the expression for $\Pi^0_z$, the simple change of variables $p_0-a_0\to p'_0$. Afterwards, one arrives at the integral 
$$
\int\frac{dp'_0d^3\vec{p}}{(2\pi)^4}\frac{p'_0}
{(p'_0)^2-(\vec{p}-\vec{a})^2(\vec{p}^2)^{2n}-m^{2z}}, 
$$
which clearly vanishes by symmetry reasons (here $\vec{p}^2=p_ip_i$ is the usual Euclidean vector square). As a result, we immediately see that $\Pi^0_z=0$. Therefore, the effective potential does not depend on $a_0$.

In the expression for $\Pi^i_z$, the denominator depends on $p_0$ in a similar way, but the numerator does not depend on $p_0$. So, we find
\begin{eqnarray}
\Pi^i_z &=& -2\delta^{ij} \int\frac{d^3\vec{p}}{(2\pi)^3}\frac{(p_j-a_j)(\vec{p}^2)^{2n}}
{[(\vec{p}-\vec{a})^2(\vec{p}^2)^{2n}+m^{2z}]^{1/2}} ,
\end{eqnarray}
However, this integral apparently cannot be evaluated exactly for the arbitrary $n$ while $a_i\neq 0$. So, the complete dependence of the effective potential on $a_i$ cannot be found explicitly, and must be obtained order by order. To do it, we use the expansion of the denominator of (\ref{pi}) in field expectations $a_i$.
The basic formula for this expansion is}
\begin{equation}
\frac{1}{\slashed{p}_0+\slashed{p}_i(p_jp^j)^n-\slashed{a}_0-\slashed{a}_i(p_jp^j)^n-m^z} = S(p) + S(p)(\slashed{a}_0+\slashed{a}_i(p_jp^j)^n)S(p)+\cdots,
\end{equation}
where $S(p)=(\slashed{p}_0+\slashed{p}_i(p_jp^j)^n-m^z)^{-1}$. Then, as we already noted, $\Pi^0_z=0$ and 
\begin{equation}
\Pi^i_z =-i\alpha_1 a^i+i\beta a^i a_ja^j+\cdots,  
\end{equation}
up to third order in $a_i$, with
\begin{equation}
\alpha_1 = -\frac{2^{-d}\pi ^{-\frac{d}{2}-\frac{1}{2}} \Gamma \left(\frac{d-2}{2 z}+1\right) \Gamma \left(-\frac{d+z-2}{2 z}\right)}{z^2 \Gamma \left(\frac{d}{2}+1\right)}(d-2)(z-1) m^{d+z-2} 
\end{equation}
and
\begin{equation}\label{beta}
\beta = \frac{2^{-d-1}\pi ^{-\frac{d}{2}-\frac{1}{2}}\Gamma \left(\frac{d-4}{2 z}+2\right) \Gamma \left(-\frac{d+z-4}{2 z}\right)}{z^3 \Gamma \left(\frac{d}{2}+2\right)}(d-4)(z-1)(d (z-1)-2 z+4) m^{d+z-4}.
\end{equation}
With this, the gap equation looks like
\begin{eqnarray}\label{gap1}
\frac{\partial V_{\mathrm{eff}}}{\partial A_0}\Big|_{A_\mu=\frac{a_\mu}{e}}&=&-\frac{1}{G_t}ea^0=0, \\
\frac{\partial V_{\mathrm{eff}}}{\partial A_i}\Bigr|_{A_\mu=\frac{a_\mu}{e}}&=&\left(-\frac{1}{G_s}-\alpha_1+\beta a_ja^j\right)ea^i=0,
\end{eqnarray}
so that $a_0=0$ and $a_ja^j=\frac{1}{\beta}\left(\frac{1}{G_s}+\alpha_1\right)$.
This confirms our conclusion that the effective potential does not depend on $A_0$. We will concentrate now on the dependence of the effective potential on $A_i$.

Let us now study the effective action. For this, we can rewrite (\ref{Seff}) as
\begin{equation}\label{acaoefetiva}
S_\mathrm{eff}[A] = \int d^4x\, \left(\frac{g_t^2}{2}A_0A^0+\frac{g_s^2}{2}A_iA^i\right)+S_{\mathrm{eff}}^{(l)}[A],
\end{equation}
with
\begin{eqnarray}\label{acao_l}
S_{\mathrm{eff}}^{(l)}[A]&=&i\mathrm{Tr}\sum_{l=1}^{\infty}\frac{1}{l}\left[S(p)e(\slashed{A}_0+\slashed{A}_i(p_jp^j)^n)\right]^l,
\end{eqnarray}
where we have disregarded $-i\mathrm{Tr}\ln(\slashed{p}_0+\slashed{p}_i(p_jp^j)^n-m^z)$, since it is field independent. For $l=1$ and $l=3$, trivially, $S^{(3)}_\mathrm{eff}[A]$ and $S^{(1)}_\mathrm{eff}[A]$ vanish, since the trace of odd number of Dirac matrices is always zero. Then, let us focus our attention on contributions $l=2$ and $l=4$, whose analysis is sufficient for making a conclusion about the possible spontaneous breaking of the rotational symmetry (see also \cite{Assuncao}, where this methodology has been applied for a dynamical Lorentz symmetry breaking). 

For $l=2$, we have
\begin{eqnarray}
S^{(2)}_\mathrm{eff}[A] &=& \frac i2 \mathrm{Tr}\,S(p)e(\slashed{A}_0+\slashed{A}_i(p_jp^j)^n)S(p)e(\slashed{A}_0+\slashed{A}_i(p_jp^j)^n) = \frac{ie^2}{2}\int d^4x \Pi_z^{\mu\nu}A_\mu A_\nu,
\end{eqnarray}
where 
\begin{equation}
\Pi_z^{\mu\nu} = \mathrm{tr} \int \frac{d^4p}{(2\pi)^4} S(p) \Gamma^\mu(p) S(p-i\partial) \Gamma^\nu(p-i\partial),
\end{equation}
with $\Gamma^\mu(p)=(\gamma^0,\gamma^i(p_jp^j)^n)$. Therefore, since $S^{(2)}_\mathrm{eff}=\int d^4x {\cal L}^{(2)}_\mathrm{eff}$, we obtain
\begin{eqnarray}\label{Leff2}
{\cal L}^{(2)}_\mathrm{eff} &=& \frac{e^2}{2}\alpha_1 A_iA^i -\frac{e^2}{2}(\alpha_2\partial_0A_i\partial^0A^i-\alpha_3\partial_0A_i\partial^iA^0-\alpha_3\partial_iA_0\partial^0A^i+\alpha_4\partial_iA_0\partial^iA^0) \nonumber\\
&&-\frac{e^2}{2}(\alpha_5\partial_iA_j\partial^iA^j-\alpha_6\partial_iA_j\partial^jA^i) + {\cal O}(\partial^4),
\end{eqnarray}
where
\begin{subequations}\label{alphas}
\begin{eqnarray}
\alpha_2 &=& \frac{2^{-d} \pi ^{-\frac{d}{2}-\frac{1}{2}} \Gamma \left(\frac{d-2}{2 z}+1\right) \Gamma \left(\frac{-d+z+2}{2 z}\right)}{3 d z^2 \Gamma \left(\frac{d}{2}\right)}(d (3 z-1)-2 z+2) m^{d-z-2}, \\
\alpha_3 &=& \frac{2^{-d} \pi ^{-\frac{d}{2}-\frac{1}{2}}\Gamma \left(\frac{d-2}{2 z}+1\right) \Gamma \left(\frac{-d+z+2}{2 z}\right)}{3 d z \Gamma \left(\frac{d}{2}\right)} (2 d+z-1) m^{d-z-2},  \\
\alpha_4 &=& \frac{2^{-d} \pi ^{-\frac{d}{2}-\frac{1}{2}} \Gamma \left(\frac{d-2}{2 z}+1\right) \Gamma \left(\frac{-d+z+2}{2 z}\right)}{3 d z \Gamma \left(\frac{d}{2}\right)}((z-1) (z+3)-d (z-3)) m^{d-z-2},
\end{eqnarray}
and
\begin{eqnarray}
\alpha_5 &=& \frac{2^{-d-3} \pi ^{-\frac{d}{2}-\frac{1}{2}} \Gamma \left(\frac{d-4}{2 z}+1\right) \Gamma \left(-\frac{d+z-4}{2 z}\right)}{3 z^2 \Gamma \left(\frac{d}{2}+2\right)} \nonumber\\
&&\times \left(-(d-4) (3 d+2) z^2+12 (d-2) d z+((2 d-9) d+6) d-8\right) m^{d+z-4}, \\
\alpha_6 &=& -\frac{2^{-d} \pi ^{-\frac{d}{2}-\frac{1}{2}}\Gamma \left(\frac{d-4}{2 z}+2\right) \Gamma \left(\frac{-d+z+4}{2 z}\right)}{3 (d+z-4) \Gamma \left(\frac{d}{2}+2\right)}(d (d-2 z+4)+8 (z-1)) m^{d+z-4}.
\end{eqnarray}
\end{subequations}
We can easily observe that, for $z=1$, $\alpha_1=0$ and $\alpha_i=\frac{1}{6\pi^2\epsilon'}$, where $\frac{1}{\epsilon'}=\frac{1}{\epsilon}-\ln\frac{m}{\mu'}$, with $\epsilon=3-d$ and $\mu'^2=4\pi e^{-\gamma}\mu^2$, and $i$ runs from $2$ to $6$, so that we have
\begin{equation}
{\cal L}^{(2)}_\mathrm{eff} = -\frac{1}{4}\frac{e^2}{6\pi^2\epsilon'}F_{\mu\nu}F^{\mu\nu},
\end{equation}
which is the result discussed in \cite{Bjorken} and confirms our expressions (\ref{alphas}). We note that we have ignored the higher derivative terms. 

Following, for the next critical point $z=3$, in the three-dimensional space, one has $\alpha_5=\alpha_6$, which allows for arising $F_{ij}F_{ij}$ term (cf. \cite{ourz3}). The case $z=3$ is of special importance because namely at $z=3$ the most important quantum corrections in Lifshitz spinor QED, such as the kinetic term for the gauge field \cite{Gomes:2018umk} and Adler-Bell-Jackiw anomaly \cite{Bakas}, differ from zero, while at $z=2$ they vanish \cite{ourHLED,ourz2}.

Now, for $l=4$, we get
\begin{eqnarray}
S^{(4)}_\mathrm{eff}[A] &=& \frac i4 \mathrm{Tr}\,S(p)e[\slashed{A}_0+\slashed{A}_i(p_jp^j)^n]S(p)e[\slashed{A}_0+\slashed{A}_i(p_jp^j)^n]S(p) \nonumber\\
&&\times S(p)e[\slashed{A}_0+\slashed{A}_i(p_jp^j)^n]S(p)e[\slashed{A}_0+\slashed{A}_i(p_jp^j)^n] \nonumber\\
&=& \frac{ie^2}{4}\int d^4 \Pi_z^{\kappa\lambda\mu\nu}A_\kappa A_\lambda A_\mu A_\nu,
\end{eqnarray}
where
\begin{equation}
\Pi_z^{\kappa\lambda\mu\nu} = \mathrm{tr} \int \frac{d^4p}{(2\pi)^4} S(p) \Gamma^\kappa(p)S(p)\Gamma^\lambda(p)\Gamma^\mu(p)S(p)\Gamma^\nu(p)+{\cal O}(\partial^4).
\end{equation}
Then, we obtain
\begin{equation}\label{Leff4}
{\cal L}^{(4)}_\mathrm{eff} =  -\frac{e^4}{4}\beta A_iA^i A_jA^j + {\cal O}(\partial^4),
\end{equation}
where $\beta$ is the same result (\ref{beta}) found earlier within the analysis of gap equation.

Therefore, we conclude that the effective potential, defined as the effective Lagrangian with an inverse sign, evaluated when derivatives of fields are equal to zero, is
\begin{equation}\label{Veff}
V_\mathrm{eff}=-\frac{g_t^2}{2}A_0A^0-\frac{g_s^2}{2}A_iA^i-\frac{e^2}{2}\alpha_1 A_i A^i+\frac{e^4}{4}\beta (A_iA^i)^2+\cdots,
\end{equation}
where dots are for sextuple and higher terms which are irrelevant in the weak field approximation and can be neglected. Now, using the gap equation (\ref{gap1}), i.e., the fact that $\frac{1}{G_s}=-\alpha_1+\beta a_ia^i$, we can rewrite (\ref{Veff}) as
\begin{equation}
V_\mathrm{eff}=-\frac{g_t^2}{2}A_0A^0+\frac{\beta}{4}(e^2A_iA^i-a_ia^i)^2,
\end{equation}
where we have added the constant $\frac{\beta}{4}(a_ia^i)^2$. Since $A_i$ is real, $A_iA^i<0$ in our signature $(+---)$. Thus, we see that the spontaneous breaking of the rotational symmetry occurs if $\beta>0$ when the effective potential possesses minima. This condition implies in a combination of inequalities $z+4-d>0$ and $(d-2)z-d+4>0$. It is easy to see that for $d=3$ these inequalities are satisfied for any (non-negative) $z$, so, the dynamical breaking of rotational symmetry occurs for any $z$.

Now, with using the expressions (\ref{acaoefetiva}), (\ref{Leff2}), and (\ref{Leff4}), for $z=3$, the effective Lagrangian can be written as
\begin{eqnarray}\label{LeffT}
{\cal L}_\mathrm{eff} &=& \frac{g_t^2}{2}A_0A^0 -\frac{\tilde\alpha_2}{2} \left(\partial_0A_i\partial^0A^i-\frac65\partial_0A_i\partial^iA^0-\frac65\partial_iA_0\partial^0A^i+\frac{36}{25}\partial_iA_0\partial^iA^0 \right) \nonumber\\
&&-\frac{\tilde\alpha_5m^4}{2}(\partial_iA_j\partial^iA^j-\partial_iA_j\partial^jA^i)-\frac{9\tilde\alpha_2}{50}\partial_iA_0\partial^iA^0 -\frac{\tilde\beta m^4e^2}{4}\left(A_iA^i-\frac{a_ia^i}{e^2}\right)^2,
\end{eqnarray}
where $\tilde\alpha_2=e^2\alpha_2$, $\tilde\alpha_5=\frac{e^2}{m^4}\alpha_5$, and $\tilde\beta=\frac{e^2}{m^4}\beta$, with
\begin{eqnarray}
\alpha_2 &=& \frac{5\Gamma \left(\frac{1}{3}\right) \Gamma \left(\frac{7}{6}\right)}{81 \pi ^{5/2} m^2}, \\
\alpha_5 &=& \frac{19 m^2 \Gamma \left(-\frac{1}{3}\right) \Gamma \left(\frac{5}{6}\right)}{324 \pi ^{5/2}}, \\
\beta &=& -\frac{4 m^2 \Gamma \left(-\frac{1}{3}\right) \Gamma \left(\frac{11}{6}\right)}{405 \pi ^{5/2}}.
\end{eqnarray}
Thus, as the mass dimensions are $[m]=1$, $[e]=1$, $[A_0]=2$, $[A_i]=0$, $[\partial_0]=3$, and $[\partial_i]=1$, obviously, we have $[\tilde\alpha_2]=0=[\tilde\alpha_5]=[\tilde\beta]$, i.e., they are dimensionless. With this, as a first attempt, we consider the rescaling 
\begin{subequations}\label{reescal}
\begin{eqnarray}
A_0 &\to& \frac{5m\tilde\alpha_5^{1/4}}{6\tilde\alpha_2^{1/2}}A_0, \\
A_i &\to& \frac{1}{m\tilde\alpha_5^{1/4}}A_i, \\
\partial_0 &\to& \frac{m^2\tilde\alpha_5^{1/4}}{\tilde\alpha_2^{1/2}}\partial_0, \\
\partial_i &\to& \frac{1}{\tilde\alpha_5^{1/4}}\partial_i,
\end{eqnarray}
\end{subequations}
$e\to m\,e$, and $G_t\to m^2G_t$, so that we obtain
\begin{eqnarray}
\label{actresc}
{\cal L}_\mathrm{eff} &=& -\frac{m^2}{4}F_{\mu\nu}F^{\mu\nu} -\frac{m^2}{8}(\partial_iA_0)^2 +\frac{g_t^2}{2}\frac{25m^2\tilde\alpha_5^{1/2}}{36\tilde\alpha_2}A_0A^0 -\frac{\tilde\beta m^2e^2}{4\tilde\alpha_5}\left(A_iA^i-\frac{a_ia^i}{e^2}\right)^2.
\end{eqnarray}
This action involves the Maxwell term, a second-derivative term $(\partial_iA_0)^2$, a mass term for $A_0$, and a bumblebee potential for $A_i$. To eliminate the mass term, we can consider our theory in the limit $g_t \to 0$, that is, the strong coupling limit, in terms of $G_t$. Thus, we get
\begin{eqnarray}
{\cal L}_\mathrm{eff} &=& -\frac{m^2}{4}F_{\mu\nu}F^{\mu\nu} -\frac{\tilde\beta m^2e^2}{4\tilde\alpha_5}\left(A_i A^i-\frac{a_i a^i}{e^2}\right)^2,
\end{eqnarray}
where we have chosen the gauge $(\partial_iA_0)^2=0$, which is allowed since $A_0$ enters only the Maxwell term and not the potential term. However, we must note that $\tilde{\alpha}_5<0$. Therefore, in favor of restoring the Maxwell term, the spontaneous symmetry breaking is lost under this rescaling. Moreover, since $\tilde{\alpha}_5<0$, the transformation (\ref{reescal}) is actually imaginary. 

On the other hand, we can consider $\tilde\alpha_5=-|\tilde\alpha_5|$ in (\ref{LeffT}), and rewrite (\ref{reescal}) in the function of $|\tilde\alpha_5|^{1/4}$, so that we obtain
\begin{eqnarray}
\label{result}
{\cal L}_\mathrm{eff} &=& -\frac{m^2}{2}F_{0i}F^{0i}+\frac{m^2}{4}F_{ij}F^{ij} -\frac{\tilde\beta m^2e^2}{4|\tilde\alpha_5|}\left(A_i A^i-\frac{a_i a^i}{e^2}\right)^2.
\end{eqnarray}
Thus, now, the spontaneous symmetry breaking indeed can occur, but the Lorentz symmetry cannot be perturbatively restored. We note that in this case we effectively start with the globally $U(1)$ symmetric Lagrangian (\ref{L0}) and arrive at the Lagrangian (\ref{result}), which turns out to possess a local $U(1)$ symmetry at minima of the potential, as occurs in the Lorentz invariant four-fermion theory \cite{Eguchi}. Indeed, under the $U(1)$ transformation, the variation of the Lagrangian (\ref{result}) is completely given by the variation of its potential, hence, if we have a gauge transformation from one minima to another, the potential is invariant. The similar situation takes place also in Lorentz-invariant theories \cite{Chkareuli}.

Therefore, we see that the resulting one-loop effective Lagrangian (\ref{result}) involves, first, the Maxwell-like kinetic term with nonconventional relative sign of mixed and purely spatial contributions, second, the bumblebee potential, for the $A_i$ fields, possessing well-defined minima and hence allowing for a spontaneous breaking of the rotational symmetry. Moreover, within the potential term, the $A_0$ and $A_i$ fields are essentially separated. The strong asymmetry between space and time directions, displayed by our one-loop quantum correction (\ref{result}), shows that there is no low-energy restoring of the Lorentz symmetry in our theory. 

Let us discuss our results. In this paper, we formulated a new four-fermion Lifshitz model whose fermion self-coupling vertex involves derivatives. For this theory, we  introduced an auxiliary vector field and explicitly obtained its one-loop effective Lagrangian  given by a sum of the Maxwell-like kinetic term and the bumblebee potential.

For specific relations between $d$ and $z$, the bumblebee potential generated by us turns out to possess a continuous set of minima, just as occurs in the usual Lorentz-breaking case in \cite{Assuncao}. Therefore, any special value of $A_i=\langle A_i \rangle$, satisfying the condition $\langle A_iA^i \rangle=\frac{1}{e^2\beta}\left(\frac{1}{G_s}+\alpha_1\right)$, corresponds to choice of one of the vacua, introducing the privileged direction in the space (but not space-time as occurs in usual Lorentz-breaking theories, remind that the Lorentz symmetry is already strongly broken since our theory is Lifshitz-like), and thus breaking the spatial rotational symmetry. Therefore, we demonstrated, for the first time, the dynamical breaking of rotational symmetry in Lifshitz-like theories. To conclude, we showed that the bumblebee methodology can be applied to Lifshitz-like theories as well. We expect that this approach can be generalized to more complicated theories, for example, those ones including the coupling of the vector and spinor fields with gravity. Also, we note that since one of the main motivations for Lifshitz-like theories consists in their possible application to describing phase transitions and critical behavior (it is worth to mention that the seminal paper \cite{Lifshitz}, where the space-time anisotropy has been originally introduced is aimed to study of phase transitions; among modern applications of Lifshitz-like theories to study of various critical phenomena, the papers \cite{Leite} deserve to be mentioned), it is natural to expect that our results can be applied as well to studies of condensed matter where spontaneous symmetry breaking could be very important.

{\bf Acknowledgements.} This work was partially supported by Conselho
Nacional de Desenvolvimento Cient\'{\i}fico e Tecnol\'{o}gico (CNPq).  The work by A. Yu. P. has been partially supported by the CNPq project 303783/2015-0.

\end{document}